\documentclass[twocolumn,secnumarabic,amssymb,nobibnotes,aps,prd,superscriptaddress]{revtex4-2}

\usepackage{siunitx}
\usepackage{booktabs}
\usepackage{amsmath}
\usepackage{commath}
\usepackage{graphicx}
\usepackage{xcolor}

\renewcommand{\vec}[1]{\boldsymbol{#1}}

\begin{document}

\title{
    On the origin of noncollinear magnetization coupling across RuX layers
}%

\author{Claas Abert}%
\email[]{claas.abert@univie.ac.at}
\affiliation{Faculty of Physics, University of Vienna, Austria}
\affiliation{Research Platform MMM Mathematics-Magnetism-Materials, University of Vienna, Austria}
\author{Sabri Koraltan}%
\email[]{sabri.koraltan@univie.ac.at}
\affiliation{Faculty of Physics, University of Vienna, Austria}
\author{Florian Bruckner}%
\affiliation{Faculty of Physics, University of Vienna, Austria}
\author{Florian Slanovc}%
\affiliation{Faculty of Physics, University of Vienna, Austria}
\author{Juliana Lisik}%
\affiliation{Simon Fraser University, 8888 University Drive, Burnaby, British Columbia V5A 1S6, Canada}
\author{Pavlo Omelchenko}%
\affiliation{Simon Fraser University, 8888 University Drive, Burnaby, British Columbia V5A 1S6, Canada}
\author{Erol Girt}%
\affiliation{Simon Fraser University, 8888 University Drive, Burnaby, British Columbia V5A 1S6, Canada}
\author{Dieter Suess}%
\affiliation{Faculty of Physics, University of Vienna, Austria}
\affiliation{Research Platform MMM Mathematics-Magnetism-Materials, University of Vienna, Austria}
\date{\today}%
\begin{abstract}
    We present a simple atomistic model for the description of noncollinear coupling in magnetic multilayers with hybrid spacer layers made of Ru alloyed to ferromagnetic atoms such as Fe.
    In contrast to previous analytical and micromagnetic models that explain the noncollinear coupling by means of lateral fluctuations in the coupling constant, the presented model accounts for atom-atom coupling in all three spatial dimensions within the spacer layer.
    The new model is able to accurately predict the dependence of the macroscopic bilinear and biquadratic coupling constants on the spacer-layer composition and thickness, showing much better quantitative agreement than lateral-fluctuation models.
    Moreover, it predicts noncollinear coupling even for infinitely stiff ferromagnetic layers which goes beyond the predictions of previous models.
\end{abstract}
\maketitle

Magnetic multilayers build the backbone of many spintronics applications such as magnetoresistive sensors \cite{zheng2019}, magnetoresistive random access memory (MRAM) \cite{bhatti2017} or spin-torque nano-oscillators \cite{chen2016spin}.
The coupling between the magnetic layers plays a crucial role in controlling the functionality of multilayer devices.
With this regard, an important effect is the Ruderman–Kittel–Kasuya–Yosida (RKKY) coupling \cite{ruderman1954indirect} that introduces an exchange coupling between two magnetic layers separated by a nonmagnetic layer, typically made of Ru.
Depending on the thickness of the nonmagnetic spacer layer, the RKKY coupling between the magnetic layers is either of parallel or antiparallel nature and hence collinear.
While a collinear coupling mechanism is useful, e.g. for the construction of synthetic antiferromagnets, a tunable control of the coupling angle would introduce numerous advantages for the design of spintronic devices.

\begin{figure}[t]
    \centering
    \includegraphics{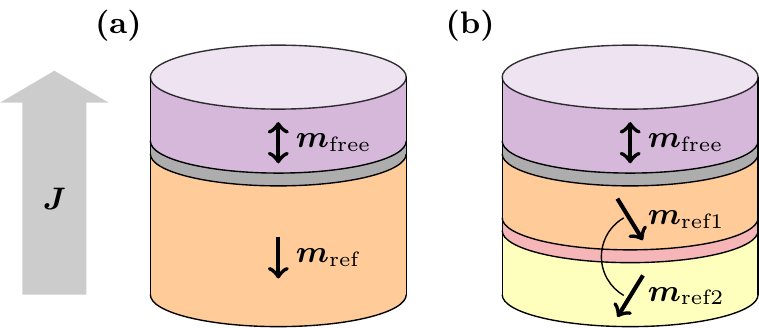}
    \caption{
        Simplified perpendicular MRAM stack with magnetic free layer $\vec{m}_\text{free}$ and magnetic reference layer $\vec{m}_\text{ref}$.
        When applying a perpendicular current, $\vec{J}$, the reference layer polarizes the itinerant electrons, resulting in a spin torque exerted on the free-layer magnetization.
        (a) Conventional collinear MRAM.
        (b) MRAM with noncollinear reference system for improved writing.
    }
    \label{fig:mram}
\end{figure}

Considering a typical spin-transfer torque MRAM device, as depicted in Fig.~\ref{fig:mram}(a), the collinear alignment of the distinct magnetic layers introduces serious drawbacks to the writing process.
Namely, the spin torque generated by a reference layer on a perfectly collinear free layer vanishes for the equilibrium configuration.
In this case, the switching of the free layer is facilitated by thermal activation.
By breaking the collinearity between the reference layer and the free layer, this restriction is overcome, allowing for a reliable switching process with low power consumption.

A possible method to avoid collinearity in spin-torque devices is the tilting of the reference layer anisotropy \cite{zhou2009a,mojumder2012a}.
Alternative strategies for noncollinear spin-polarization include the use of two reference layers, one being in-plane and the other being out-of-plane \cite{law2009a,sbiaa2016a} or the combination of an out-of-plane spin-polarization layer with an in-plane free layer \cite{kent2004a,suess2017significant}.
Besides the enhancement of MRAM performance, noncollinear magnetic multilayers have already been proven to be beneficial for spin-torque oscillators \cite{zhou2008a,skowronski2012zero,arun2020a} and are likely to play a crucial role for designing a variety of future spintronic devices.

Recently, a novel mechanism for the noncollinear coupling in magnetic multilayers has been experimentally demonstrated \cite{nunn2020control} and it is shown that spacer layers made of RuFe alloys are able to generate a strong noncollinear coupling between two Co layers.
The angle of this coupling can be precisely controlled by adjusting the ratio of the RuFe composition in the spacer layer.
This introduces a very powerful tool for the development of novel spintronic devices.
Figure~\ref{fig:mram}(b) depicts an MRAM structure with noncollinear alignment between the free and reference layers, which is achieved by using a reference layer that consists of two noncollinearly-coupled magnetic layers.
Such a reference system would allow for a tilted spin polarization in the free layer and hence lower the critical switching current \cite{zhou2009a}.
Other implications such as magnetization oscillations due to spin-torque effects within the noncollinear reference system might introduce further implications for the device optimization.

In \cite{nunn2020control}, we developed a simple micromagnetic model based on a fluctuation mechanism introduced in \cite{slonczewski1991fluctuation} that divides the spacer layer laterally into regions of ferromagnetic coupling and antiferromagnetic coupling.
In this model, the coupling energy across the RuFe spacer layer is given by the interface integral
\begin{equation}
    E = \int_{\Gamma} - A(\vec{x}) \, \vec{m_1}(\vec{x}) \cdot \vec{m_2}(\vec{x}) \dif{\vec{s}},
\end{equation}
with $\Gamma$ being the interface between the magnetic layers, $\vec{m}_1$ and $\vec{m}_2$ being the respective magnetization configurations in these layers and $A$ being the spatially-varying coupling constant.
If the spatial fluctuations in $A$ are on a length scale $L_\text{fluc}$ that is small compared to the exchange length $L_\text{ex}$ of the ferromagnetic layers, $L_\text{fluc} \ll L_\text{ex}$, the effective coupling of the magnetization in the ferromagnetic layers $\vec{m}_1$ and $\vec{m}_2$ amounts to the average coupling constant $\bar{A}$, resulting in an areal energy density
\begin{equation}
    \epsilon = - \bar{A} \, \vec{m}_1 \cdot \vec{m}_2
\end{equation}
and collinear coupling of the ferromagnetic layers.
If $L_\text{fluc}$ is large compared to $L_\text{ex}$, the ferromagnetic layers will couple region by region, leading to a domain pattern defined by the distribution of ferromagnetically ($A > 0$) and antiferromagnetically ($A < 0$) coupled regions.
However, if $L_\text{fluc} \approx L_\text{ex}$, the fluctuations in the coupling constant are able to generate slight inhomogeneities in the ferromagnetic layers without generating domains.
In this case, the magnetization in the ferromagnetic layers can be assumed to be approximately homogeneous and the coupling of the ferromagnetic layers can be described by adding an additional biquadratic term to the coupling energy,
\begin{equation}
    \epsilon = - A_1 \vec{m}_1 \cdot \vec{m}_2 - A_2 (\vec{m}_1 \cdot \vec{m}_2)^2.
\end{equation}
The values of $A_1$ and $A_2$ depend on various system parameters such as the exchange constant of the ferromagnetic layers and the exact distribution of the coupling strength $A(\vec{x})$.
For negative $A_2$ and $|A_1| < |2A_2|$, noncollinear coupling of the ferromagnetic layers becomes energetically stable.
In order to theoretically determine the macroscopic coupling constants $A_1$ and $A_2$, a microscopic model that resolves the inner structure of the spacer layer is required.
While the micromagnetic model based on lateral fluctuations in $A(\vec{x})$ has been shown to provide a possible explanation for the noncollinear coupling in Co-RuFe-Co multilayers \cite{nunn2020control}, it fails to accurately describe all experimentally observed effects of the coupling mediated by RuFe layers.
For instance, the $A_1$ as predicted by the micromagnetic model exhibits a linear dependence on the fraction of Ru content in the spacer layer while the experiment shows saturation for high Ru content \cite{nunn2020control}.
This leads to unrealistically high values of $A_1$ when fitting the model parameters to reproduce the experimentally observed $A_2$.
Moreover, the micromagnetic model has a low predictivity for the dependence of $A_1$ and $A_2$ on the spacer-layer thickness since it requires the thickness-dependent antiferromagnetic coupling constant as an input.

\begin{figure}[h]
    \centering
    \scalebox{0.8}{
    \includegraphics{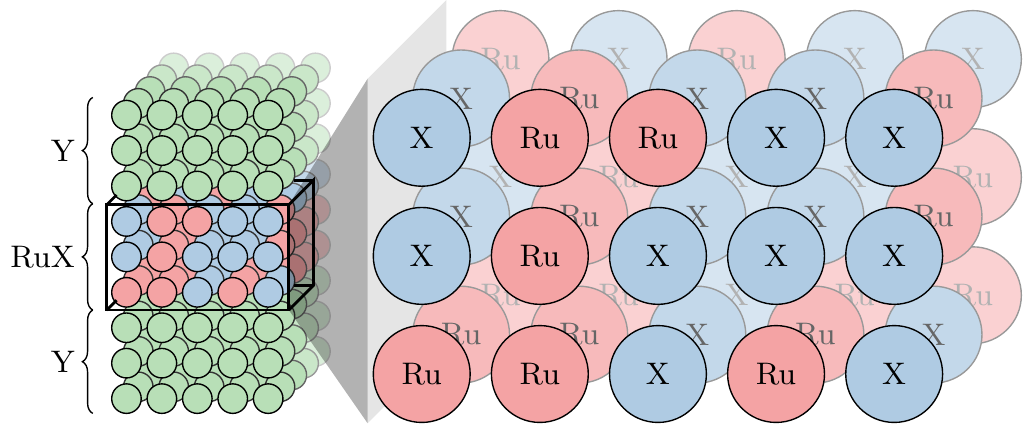}}
    \caption{
        Atomistic model of ferromagnetic multilayer structure with alloy spacer layer containing ferromagnetic (X/Y) as well as Ru atoms in a cubic lattice.
    }
    \label{fig:lattice}
\end{figure}
In order to overcome the weaknesses of the micromagnetic model, we employ a simple atomistic model that considers pair coupling of neighboring ferromagnetic atoms as well as indirect coupling of ferromagnetic atoms across Ru atoms.
For the sake of simplicity, we assume a cubic lattice where each atomic site is either populated by a ferromagnetic atom X/Y or a Ru atom, see Fig.~\ref{fig:lattice}.
\begin{table}
    \caption{
        Exemplary couplings of ferromagnetic atoms depending on neighboring Ru atoms.
    }
    \label{tab:model}
    \begin{tabular}{ll}
        \toprule
        \textbf{Type} & \textbf{Energy}\\ \midrule

        \raisebox{-.5\height}{\includegraphics{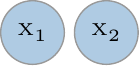}} & $E = - J^\text{X/X}_1 \vec{m}_{\text{X}_1} \cdot \vec{m}_{\text{X}_2}$ \\ \midrule
        \raisebox{-.5\height}{\includegraphics{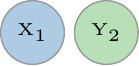}} & $E = - J^\text{X/Y}_1 \vec{m}_{\text{X}_1} \cdot \vec{m}_{\text{Y}_2}$ \\ \midrule
        \raisebox{-.5\height}{\includegraphics{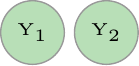}} & $E = - J^\text{Y/Y}_1 \vec{m}_{\text{Y}_1} \cdot \vec{m}_{\text{Y}_2}$ \\ \midrule
        \raisebox{-.5\height}{\includegraphics{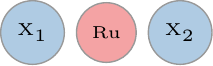}} & $E = - J^\text{X/Ru/X}_2 \vec{m}_{\text{X}_1} \cdot \vec{m}_{\text{X}_2}$ \\ \midrule
        \dots & \\ \midrule
        \raisebox{-.5\height}{\includegraphics{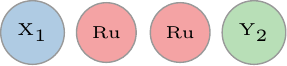}} & $E = - J^\text{X/Ru/Y}_3 \vec{m}_{\text{X}_1} \cdot \vec{m}_{\text{Y}_2}$ \\ \midrule
        \dots & \\ \bottomrule
    \end{tabular}
\end{table}
For neighboring ferromagnetic atoms in the principal directions of the cubic lattice, we employ the classical Heisenberg model
\begin{equation}
    \mathcal{H} = - J \vec{S}_1 \cdot \vec{S}_2,
\end{equation}
with $J$ being the coupling constant and $\vec{S}_1$ and $\vec{S}_2$ being unit vectors that represent the coupled spins.
In order to account for the influence of the Ru atoms in the spacer layer, we assume that the ferromagnetic atoms are also coupled when separated by one or more Ru atoms in one of the three principal directions of the lattice, see Tab.~\ref{tab:model}.
The Hamiltonian of our atomistic model is completed by the Zeeman energy and a uniaxial anisotropy that accounts for both crystalline as well as shape anisotropy, resulting in 
\begin{equation}
    \mathcal{H} =
    - \sum_{i,j} J_{k(i,j)} \vec{S}_i \cdot \vec{S}_j
    - \sum_{i}
    \mu_0 \mu_i \vec{H} \cdot \vec{S}_i
    + K a^3 (\vec{S}_i \cdot \vec{e}_k)^2,
    \label{eq:energy}
\end{equation}
with $J_{k(i,j)}$ being the coupling constant according to Tab.~\ref{tab:model}, $\mu_i$ being the magnetic moment of a single atom, $K$ being the anisotropy constant and $\vec{e}_k$ being the unit vector along the uniaxial anisotropy axis.
The coupling constants $J_k$ for the atomistic model can be determined by macroscopic considerations.
The nearest-neighbor coupling $J_1$ is chosen as $J_1 = A_\text{ex} a$ with $A_\text{ex}$ being the exchange constant and $a$ being the (artificial) cubic lattice constant in order to reproduce accurate ferromagnetic behavior in homogeneous layers.
The coupling constant $J_k$ with $k > 1$ can be chosen according to the areal interlayer exchange coupling strength of multilayers with pure Ru spacer layers, $A(d)$, with layer thickness $d$ as $J_k = A(k a) a^2$ \cite{mckinnon2021spacer}.
In this way, the atomistic model is expected to perfectly reproduce micromagnetic models of single ferromagnetic layers as well as coupling of magnetic layers across a pure Ru spacer layer.

In order to find stable magnetization configurations for arbitrary spacer layer compositions, we minimize \eqref{eq:energy} with respect to the spin configuration $\vec{S}_i$, considering the unit sphere constraint $|\vec{S}_i| = 1$.
With this minimization procedure, the equilibrium magnetization configuration and the angle between the ferromagnetic layers can be obtained.
In order to determine the macroscopic coupling constants $A_1$ and $A_2$ for a specific spacer-layer configuration, the equilibrium spin configuration for various in-plane external fields $H$ is determined for a symmetric system with two identical ferromagnetic layers with an effective easy-plane anisotropy.
In this configuration, the magnetization in the ferromagnetic layers can be considered symmetric around the field direction.
This means that $\vec{m}_1 \cdot \vec{e}_\text{field} = \vec{m}_2 \cdot \vec{e}_\text{field} = \cos(\theta/2)$ with $\vec{e}_\text{field}$ being the direction of the in-plane external field and $\theta$ being the angle between the macroscopic magnetizations $\cos(\theta) = \vec{m}_1 \cdot \vec{m}_2$.
With this choice of field and anisotropy, the areal energy density of the system only depends on the angle $\theta$ and reads
\begin{multline}
    \epsilon(\theta) = 
    - A_1 \cos(\theta) - A_2 \cos^2(\theta) \\
    - 2 d_\text{fm} \left[
        H \mu_0 M_s \cos(\theta / 2)
        + K \sin^2(\theta / 2)
    \right],
\end{multline}
with $d_\text{fm}$ being the thickness of the ferromagnetic films.
The macrosopic coupling constants $A_1$ and $A_2$ can then be determined from atomistic simulations by fitting simulated values of $\theta$ to the equilibrium energy density condition $\dif\epsilon/\dif\theta(H,A_1,A_2) = 0$.

\begin{figure}
    \includegraphics{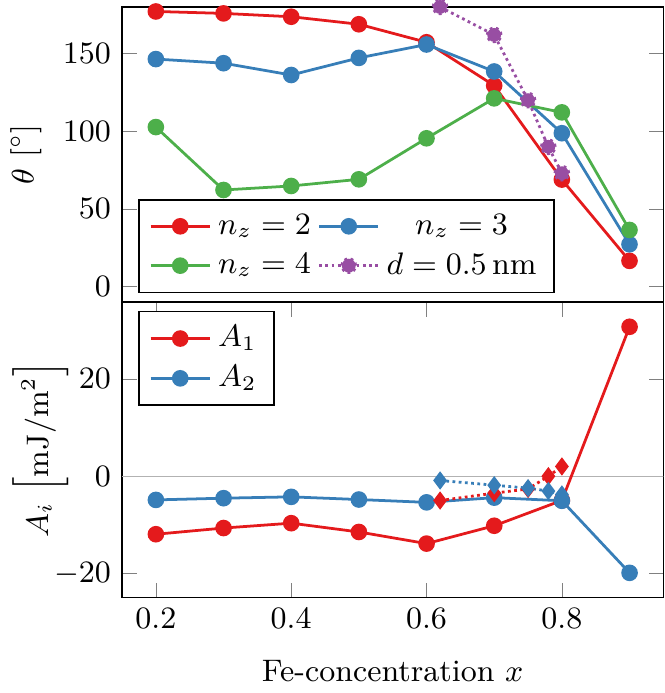}
    \caption{
        Equilibrium angle $\theta$ and macroscopic coupling constants $A_1$ and $A_2$ for a Co-RuFe-Co multilayer system for various Fe concentrations $x$ in the spacer layer.
        Simulations for $\theta$ were performed for various spacer-layer thicknesses given in atomic layers $n_z$.
        Simulations for $A_1$ and $A_2$ were performed for $n_z = 3$.
        The dotted lines mark experimental values for a spacer-layer thickness of $d = \SI{0.5}{nm}$, see \cite{nunn2020control}.
    }
    \label{fig:theta_j1j2}
\end{figure}
The atomistic model is benchmarked against the experimental findings for the Co-RuFe-Co multilayer introduced in \cite{nunn2020control}.
As a lattice constant, we choose $a = \SI{0.23}{nm}$ throughout the complete system.
In order to accurately account for the exchange coupling within the ferromagnetic layers, we compute the Heisenberg exchange constants $J_1$ from the respective exchange constants $A$ as $J_1 = A a$.
Namely, we use $J^\text{Co/Co}_1 = \SI{6e-21}{J}$ and $J^\text{Co/Fe}_1 = J^\text{Fe/Fe}_1 = \SI{4.5e-21}{J}$.
The dipole moments of our model $\mu$ are obtained from the saturation magnetization $M_s$ as $\mu = M_s a^3$.
For the sake of simplicity, we set $\mu^\text{Co} = \mu^\text{Fe} = 2.11 \mu_B$.
In order to account for the dipole--dipole interaction, we introduce an effective anisotropy in the Co layers that accounts for both, the crystalline as well as the shape anisotropy, $K_\text{eff} = K_\text{cryst} - \mu_0 M_s^2 / 2 = \SI{-7.1e5}{J/m^3}$, see \cite{nunn2020control}.
For the (antiferromagnetic) coupling across Ru atoms, we choose $J^\text{Fe/Ru/Fe}_k = - \SI{1.85e-40}{Jm^2} / (ka)^2$, $J^\text{Co/Ru/Fe}_k = - \SI{2.38e-40}{Jm^2} / (ka)^2$ and $J^\text{Co/Ru/Co}_k = - \SI{7.94e-40}{Jm^2} / (ka)^2$.
We consider only couplings across up to 4 Ru atoms which justifies the simplified dependence on the distance $ka$.
While the coupling mediated by Ru atoms is usually expected to oscillate and change sign with the distance, a purely antiferromagnetic coupling with a decay of $1/(ka)^2$ is in agreement with the experimental data on Co/Ru/Co multilayers reported in \cite{nunn2020control} and has proven to result in good macroscopic results.
The distribution of ferromagnetic and nonmagnetic atoms in the spacer layer is randomly generated according to the respective composition.
In order to find stable magnetization configurations, we use a random configuration as an initial value and apply an adaptive steepest-descent minimizer to the energy functional \eqref{eq:energy}.

Figure~\ref{fig:theta_j1j2} shows the simulation results for a system with $40\times40$ lateral spins and ferromagnetic layer thicknesses of 10 atomic layers each.
The equilibrium angle $\theta$ shows a decent agreement to the experimental data.
However, for $n_z = 4$ the simulated equilibrium angle $\theta$ exhibits a notable dip around $x = 0.4$ that is not seen in experiment.
We account this to the simplifications of our model such as the assumption of a cubic lattice.
The simulated values of $A_1$ and $A_2$ show a good agreement specifically with respect to the trends of the experimental findings, such as the the nonlinearity of $A_1(x)$, which cannot be explained by the micromagnetic model.
\begin{figure}
    \includegraphics{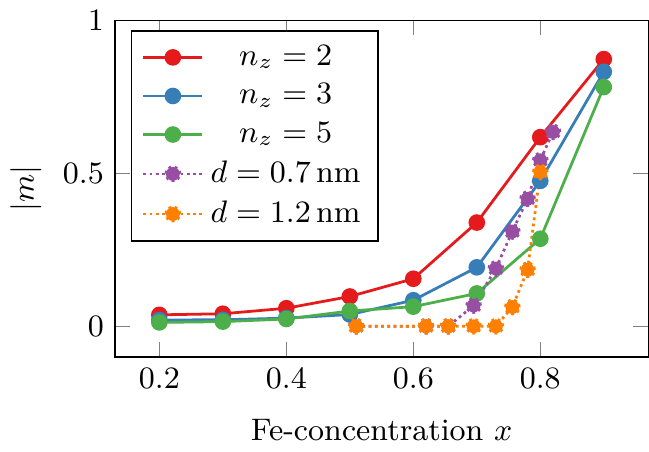}
    \caption{
        Relative saturation of spacer-layer magnetization (1 means full Fe-saturation) depending on Fe-concentration and spacer layer thickness.
        Solid lines represent simulation results with layer thicknesses given in number of atomic layers $n_z$ and dotted lines represent experimental data with layer thicknesses $d$.
    }
    \label{fig:x_vs_m}
\end{figure}
In another series of simulations, we investigate the total magnetic moment of the spacer layer for various spacer-layer thicknesses and Fe concentrations, see Fig.~\ref{fig:x_vs_m}.
The simulations accurately reproduce the experimental trend of steeper ascent of the saturation magnetization for thicker spacer layers.

\begin{figure}
    \includegraphics{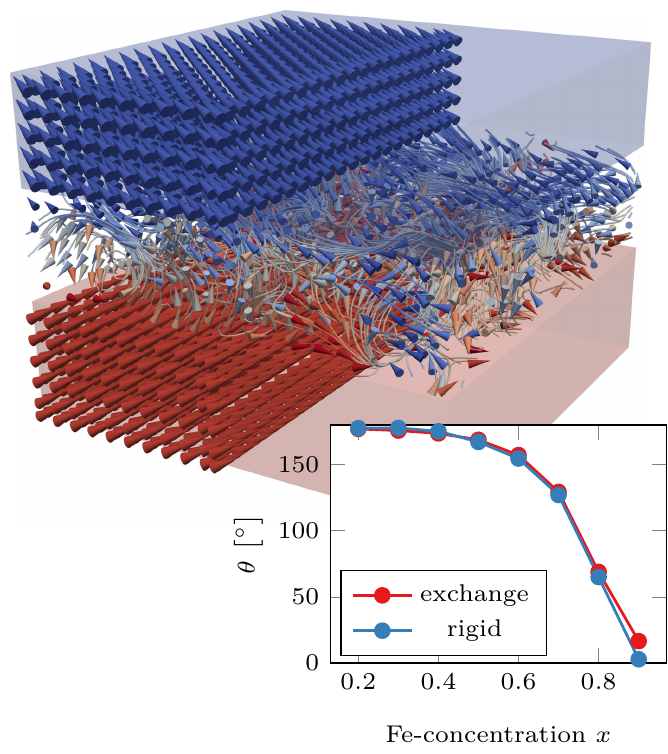}
    \caption{
        Noncollinear magnetization configuration of a magnetic multilayer with rigid ferromagnetic layers and spacer-layer thickness of 5 atomic layers.
        The plot shows the simulated equilibrium magnetization angle for various Fe concentration for ferromagnetic layers with realistic exchange coupling (exchange) compared to infinitely stiff ferromagnetic layers (rigid) for a spacer-layer thickness of 2 atomic layers.
    }
    \label{fig:mag_rigid}
\end{figure}
In order to understand this behavior, which is not expected from the micromagnetic model, the magnetization configuration is investigated in detail.
Figure~\ref{fig:mag_rigid} depicts the equilibrium configuration for a multilayer with a spacer-layer thickness of 5 atomic layers.
As indicated by the streamlines, the spacer-layer magnetization exhibits a nontrivial magnetization configuration with multiple vortex-like structures.
These vortices are obviously caused by the complicated interplay of couplings between the ferromagnetic atoms within the spacer layer, including ferromagnetic and antiferromagnetic couplings.
Note that the spacer layer itself exhibits a net magnetization at its top interface that is tilted compared to its bottom interface and therefore mediates the noncollinear coupling between the top and bottom ferromagnetic layers.
This kind of coupling mechanism is fundamentally different from the micromagnetic model proposed by Slonczewski and used in our former work.
In the micromagnetic model, the ferromagnetic regions of the spacer layer are basically assumed to be rigid and the noncollinearity is a mere result of the alternating ferromagnetic and antiferromagnetic coupling that result in slight fluctuations in the magnetization configuration of the ferromagnetic layers \cite{nunn2020control}.
Hence, the stability of noncollinear states as described by the Slonczewski model highly depends on the exchange constant of the ferromagnetic layers $A$ and vanishes for infinite stiffness since $A_2 \propto 1/A$.
The proposed atomistic model, however, enables noncollinear coupling even for infinitely stiff ferromagnetic layers, since the noncollinearity evolves within the spacer layer itself, as shown in Fig.~\ref{fig:mag_rigid}.
In order to investigate the influence of the exchange constant, we compute the equilibrium angle of the magnetization for a realistic exchange stiffness within the ferromagnetic layers and compare the result with a similar three-layer structure having infinitely stiff ferromagnetic layers.
The results shown in the plot of Fig.~\ref{fig:mag_rigid} demonstrate the small influence of the exchange stiffness of the ferromagnetic layers on the simulation outcome.

In conclusion, we present a novel atomistic model for the description of magnetic multilayer structures with spacer layers made from Ru alloyed to ferromagnetic material.
Our model accounts for the influence of the Ru atoms by means of additional Heisenberg coupling terms that couple magnetic atoms separated by Ru atoms in an antiferromagnetic fashion.
We find that the proposed model is able to reproduce the experimental results to a high level of detail, which is not accomplished by the micromagnetic model of Slonczewski that was used in our former publications \cite{nunn2020control}.
Our model is able to correctly describe trends with respect to the change of composition and thickness of the spacer layer and predicts noncollinear coupling even for infinitely stiff ferromagnetic layers.
The prediction of a noncollinear coupling which is largely independent from the exchange stiffness of the ferromagnetic layers has tremendous implications on the future development of spintronic devices and needs to be validated experimentally.

\bibliographystyle{apsrev4-1}
\bibliography{refs}

\end{document}